\documentclass[prd,aps,floats,twocolumn,twoside,preprintnumbers,
superscriptaddress,floatfix,nofootinbib,showpacs]{revtex4-2}
\usepackage{graphicx,colordvi,pstricks,rotating,amsmath}
\usepackage{amssymb,epstopdf,bbm,gensymb}
\usepackage{slashed}
\usepackage[T1]{fontenc}
\usepackage{epsfig}

\begin{document}

\title {Remarks on Compton forward scattering singularity in the NCQED}

\author{Josip Trampeti\'{c}}
\affiliation{Ru\dj er Bo\v{s}kovi\'{c} Institute, Bijeni\v{c}ka 54, 10000 Zagreb, Croatia}
\email{josip.trampetic@irb.hr}
\affiliation{Max-Planck-Institut f\"ur Physik, (Werner-Heisenberg-Institut), F\"ohringer Ring 6, D-80805 M\"unchen, Germany}
\email{trampeti@mppmu.mpg.de}
\author{Jiangyang You}
\affiliation{Ru\dj er Bo\v{s}kovi\'{c} Institute, Division of Physical Chemistry, Bijeni\v{c}ka 54, 10000 Zagreb, Croatia}
\email{jiangyang.you@irb.hr}

\newcommand{\tr}{\hbox{tr}}
\def\BOX{\mathord{\vbox{\hrule\hbox{\vrule\hskip 3pt\vbox{\vskip3pt\vskip 3pt}\hskip 3pt\vrule}\hrule}\hskip 1pt}}


\date{\today}


\begin{abstract}
We investigate the behaviour of head-to-head high energy Compton scattering in the $\theta$-exact 
NCQED in the amid its unresolved forward scattering singularity. We model the effect of the currently unknown resolution to this unavoidable singularity by cut-offs which are functions of the control variable $s/\Lambda_{\rm NC}^2$, and find that such cut-off functions can suppress NC contributions to the inverse Compton scattering to negligible. Discussed cut-offs leave the unpolarised integral cross section of the forward region larger than the backward region all the time, making the former potentially visible in the future linear collider experiments when the NC scale $\Lambda_{\rm NC}\gtrsim1$ TeV.
\end{abstract} 


\maketitle
\section{Introduction}

One of the first tree level processes studied in the noncommutative (NC) quantum electrodynamics (QED) spanned on the Moyal space \cite{Filk:1996dm,Martin:1999aq,Seiberg:1999vs,Madore:2000en,Jurco:2000fb,Jackiw:2001jb,Brandt:2003fx,Trampetic:2015zma,Gomis:2000zz,Aharony:2000gz} was high energy Compton scattering ($e\gamma\to e\gamma$) \cite{Godfrey:2001yy,Mathews:2000we}. Recently, motivated by the earlier finding that the U(N) quantum noncommutative theories constructed with or without $\theta$-exact Seiberg-Witten (SW) maps are equivalent at quantum level \cite{Martin:2016saw,Martin:2016zon,Martin:2016hji}, the Compton scattering  \cite{Godfrey:2001yy,Mathews:2000we} and other tree level NCQED scatterings \cite{Hewett:2000zp} were restudied \cite{Horvat:2020ycy,Latas:2020nji} and shown to be invariant under reversible $\theta$-exact SW map \cite{Latas:2020nji}. Considering NCQED quantum properties let’s stress that 1-loop 2-point function were started to be investigated at the very beginning of the NC theories development \cite{Filk:1996dm,Martin:1999aq} till nowadays, leading to discovery of celebrated UV/IR mixing effect \cite{Minwalla:1999px,Matusis:2000jf,VanRaamsdonk:2000rr,Hayakawa:1999yt,VanRaamsdonk:2001jd,Horvat:2011bs,Schupp:2008fs,Grosse:2005iz,Meljanac:2017jyk,Martin:2020ddo}.

One important property of Compton process which was not emphasised before, namely the forward scattering   (small-angle scattering) singularity,
was reported in \cite{Latas:2020nji}. In \cite{Trampetic:2021awu,Trampetic:2022tij} we have also shown that the  forward scattering singularity of the $f\gamma\to f\gamma$ scatterings in the $\theta$-exact SW map based NCQED theory with multiple generations of charged fermions ($f$) remain the same as in the $\theta$-exact non-SW mapped NCQED theory.

The existence of the forward scattering singularity, arising from the basic property of NCQED on Moyal space, breaking of Lorentz invariance leading to the breaking of angular momentum conservation and Bose statistics, known as Landau-Pomeranchuk-Yang (LPY) theorem, which allows the existence of the point-wise multi-photon couplings, suggests that our current understanding of Moyal NCQED is incomplete from tree level on. Its consequences on phenomenology may also be intriguing since this singularity takes place opposite to the commutative QED, in which the backward  scattering is enhanced up to the regulation by the electron mass \cite{Peskin:1995ev}. Assuming that an unknown regulation of the forward scattering singularity exists, one may ask whether the regulated differential cross section will have strong forward scattering enhancement, especially in the inverse Compton scattering (high energy electron/low energy photon) since that suppress the production of high energy photons by inverse Compton scattering. 

In this work we investigate further into the high energy Compton scattering in the $\theta$-exact NCQED model with one generation of left charge fermions: 
\begin{eqnarray}
S^{\rm min}&=&\int-\frac{1}{4}F^{\mu\nu}\star F_{\mu\nu}\;+\;\bar\Psi \star(i\slashed{D}-m)\Psi,
\label{NCminAction}\\
&\stackrel{\rm SW}{=}&S_{\rm U(1)}+S_{a^3}+S_{a^4}+S_{\bar\psi a\psi}+S_{\bar\psi a^2\psi}+\cdot\cdot\cdot.
\label{NCminAction2}
\end{eqnarray} 
The model action could be either without or with the SW map. The technical details of Eq.(\ref{NCminAction2}) and corresponding Feynman rules follows exactly those in~\cite{Latas:2020nji}. 

In the next section we present pure NCQED contribution to the unpolarised Compton cross section with respect to the forward singularity and estimate the bounds on the scale of noncommutativity. However note that there is an interplay of the noncommutative structure of spacetime and QED, i.e. interference of the tree-level NCQED with QED radiative corrections \cite{Denner:1998nk,Lee:2021iid}, which we also discuss. Finally in the last section we give summary and conclusions.

\section{Compton scattering in the NCQED and the forward scattering singularity}

From the above action (\ref{NCminAction2}), with various $\star$-products involving noncommutative $\theta$-parameter, being defined in \cite{Trampetic:2015zma}, we obtain relevant Feynman rules and compute unpolarised Compton,  
${\gamma(k_1){\bf e}^-(k_2)\to {\bf e}^-(k_3)\gamma(k_4)}$, cross section. Since in our computations are going to appear all combinations of NC phase factors of the type $k_i\theta k_j$, to determine them we choose the incoming  head-to-head particle momenta $k_1=(E^\gamma_{in},\bf k_1)$, and $k_2=(E^e_{in},\bf k_2)$ to lie on the $z$-axis. For outgoing scattered particles $k_3=(E^e_{out},\bf k_3)$, and $k_4=(E^\gamma_{out},\bf k_4)$ we choose spherical coordinate system, where by using energy $(E^\gamma_{in}+E^e_{in}=E^e_{out}+E^\gamma_{out})$ and 3-momenta $(\bf k_1+\bf k_2=\bf k_3+\bf k_4)$ conservations, we have trivially ${\bf k_1\cdot \bf k_4}={E_{in}^\gamma}^z {|\bf k_4|}\cos\vartheta$. Mandelstam variables read: $s=4E^\gamma_{in}E^e_{in}$, $t=-2E^\gamma_{in}E^\gamma_{out}(1-x)$, and $u=-2E^e_{in}E^\gamma_{out}(1+x)$, where $x=\cos\vartheta$.

The noncommutative contribution to the Compton cross section is valid for all NC types $\theta^{ij}=\frac{c^{ij}}{\rm \Lambda_{NC}^2}$, with coefficients $c^{ij},\;\forall i,j=0,1,2,3$, being of order one, and the $\Lambda_{\rm NC}$ $(\sim{\it distances}^{-1})$ is the NC scale. We have three types of space-time noncommutativity \cite{Latas:2020nji,Gomis:2000zz,Aharony:2000gz}: \\
(i) The pure space-space (spacelike) NC, where $c_{0i}=0$; $c_{ij}\not=0, \forall i,j=1,2,3$, makes unitary safe NC theory, \\
(ii) the pure time-space (timelike) NC, with $c_{0i}\not=0$; $c_{ij}=0, \forall i,j=1,2,3$, makes unitarity breaking theory,\\
(iii) the so called lightlike noncommutativity satisfying ${\bf E}^2_{\theta}={\bf B}^2_{\theta}$ and ${\bf E}_{\theta} \cdot {\bf B}_{\theta}=0$, where the $c_{0i}$ elements are defined by the direction of the background ${\bf E}_{\theta}=\frac{1}{\Lambda_{\rm NC}^2}(  {\rm c_{01}},  {\rm c_{02}},  {\rm c_{03}})$ field, while the  $c_{ij}$ coefficients are defined by the direction of the ${\bf B}_{\theta}=\frac{1}{\Lambda_{\rm NC}^2} ({\rm c_{23}},  -{\rm c_{13}}, {\rm c_{12}})$ field, produces unitary safe NCQED theory. 

The tree level Compton scattering in NCQED is one of the earliest studied \cite{Mathews:2000we,Godfrey:2001yy} and probably also one of the less understood process after two decades of research. Since it is quite obvious that the NC scale should lie above the SM scales ($m_W, m_H, m_t)$, it is reasonable to assume that $s\ll4\Lambda_{\rm NC}^2$. Thus in the pure NC differential unpolarised Compton cross sections from \cite{Latas:2020nji} we may expand cosine and Bessel functions over the small argument and obtain:
\begin{eqnarray}
&&\frac{d\sigma_{\rm NC}}{dx}=\frac{2\pi\alpha^2}{\Lambda^4_{\rm NC}}\frac{{E^\gamma_{in}}^4{E^e_{in}}^2}{\big(E^\gamma_{in}(1-x)+E^e_{in}(1+x)\big)^4}
\nonumber\\&&\cdot
\bigg[(1-x)+2\frac{E^e_{in}}{E^\gamma_{in}}(1+x)+2\left(\frac{{E^e_{in}}}{{E^\gamma_{in}}}\right)^2\frac{(1+x)^2}{(1-x)}\bigg]
\nonumber\\&&\cdot
\bigg[2c^2_{03}(1-x)+C^2(1+x)\bigg],
\nonumber\\
\label{TCrossSectC4}
\end{eqnarray}
where $C=\sqrt{(c_{01}-c_{13})^2+(c_{02}-c_{23})^2}$. There is unavoidable forward $(x\to1)$ scattering singularity in upolarised differential cross section (\ref{TCrossSectC4}). Recall that this singularity, arising from the t-channel NC amplitude ${\cal M}^{\rm C}_{It}$ (Fig. 2 and Eq. (46) in \cite{Latas:2020nji}), is originating from the triple photon coupling in the NCQED. This unique existence of an unregulated forward scattering divergence in this process was rarely mentioned in the literature until recently. The forward scattering singularity raises the concern whether the theory has to be significantly modified for consistency and/or easily distinguishable from the commutative theory. These questions are still unanswerable right now as the correct regulation of that divergence is still at large. On the other hand, a Laurent expansion near the pole $x=1$ shows that the singular term in the center-of-mass (CM) frame has the following form:
\begin{equation}
\frac{d\sigma_{\rm NC}}{dx}\sim \frac{\alpha^2}{{E_{in}^e}^2}\frac{s^2}{\Lambda_{\rm NC}^4}\frac{1}{1-x}.
\label{DCrossSectNC2}
\end{equation}
where the differential cross section bears a strong suppression factor $s^2/\Lambda_{\rm NC}^4$.
And, if the regulation of this singularity could be made equivalent to a reduced upper limit of cut-off $\Delta\lesssim 1$ when calculating integral cross sections, then the cut-off dependence would be a logarithm $\ln (1-\Delta)$. Such a logarithmic divergence could not compete with the suppression factor $s^2/\Lambda_{\rm NC}^4$ when $(1-\Delta)$ depends, more likely,  polynomial-wise on the same quantity. For these reasons the NC corrections to the $e\gamma\to e\gamma$ cross section shall remain small after the (currently unknown) regularization. And it will definitely be small if the control variable $\delta=s/\Lambda_{\rm NC}^2$ based cut-off $\Delta(\delta)\sim 1-\delta^n,\; n>0$ is used to estimate the NC effect. In particular, the integral cross sections of the inverse Compton scattering, where ${E_{in}^e}\ggg {E_{in}^\gamma}$, will be certainly negligible under such a cut-off.

Interference between tree-level NCQED and one-loop QED amplitudes, contributing to unpolarised Compton scattering differential cross section, in the CM frame bears the same strong suppression factor $s^2/\Lambda_{\rm NC}^4$ as (\ref{DCrossSectNC2}), but that contribution is proportional to $\sim\alpha^3$, i.e. it is additionally suppressed by $\alpha$. Also note that the $\varphi$-integration in arbitrary frame, of the $\theta$-dependent NC-phase part of the t-channel amplitude ${\cal M}^{\rm C}_{It}$ (from Eq. (46) in \cite{Latas:2020nji} for head-to-head $e\gamma\to e\gamma$ collision) produces leading terms proportional $(1-x)$, or to zero.

\section{Ratio of NCQED versus QED Compton cross sections}

Assuming that the differential cross section in the forward region $x\ge 0$ remains larger than the backward region after the currently unknown regulation, we attempt to use the integrated differential cross section within a forward cone $x\in [x_0,1]$ as the criterion for bounding the NC scale $\Lambda_{\rm NC}$. We also choose to simulate the unknown regulation of the forward scattering singularity by a hard cut-off near pole $x=1$. 

The quantity of interest is the ratio $\mathbb R$ between NCQED and commutative QED unpolarised integrated cross sections:
\begin{equation} 
\mathbb R=\frac{\sigma_{\rm NC}}{\sigma_{\rm QED}},
\label{R}
\end{equation}
where 
\begin{equation}
\sigma_{\rm NC}=\int\limits_{x_0}^{\Delta(\delta)} dx\; \frac{d\sigma_{\rm NC}}{dx},\;\;
\sigma_{\rm QED}=\int\limits_{x_0}^1 dx\; \frac{d\sigma_{\rm QED}}{dx}. 
\label{7}
\end{equation}
The QED differential cross section $d\sigma_{\rm QED}/dx$ 
is given in \cite{Peskin:1995ev}, for example. First we took the $E^\gamma_{in}=200$ GeV, $E^e_{in}=250$ GeV scenario as used in our prior work \cite{Latas:2020nji}, and consider $\Lambda_{\rm NC}$ from one to fifty TeV, then calculate the ratio $\mathbb R$ for $x_0=0$ and $x_0=0.5$. 

\begin{figure}
\begin{center}
\includegraphics[width=9cm,angle=0]{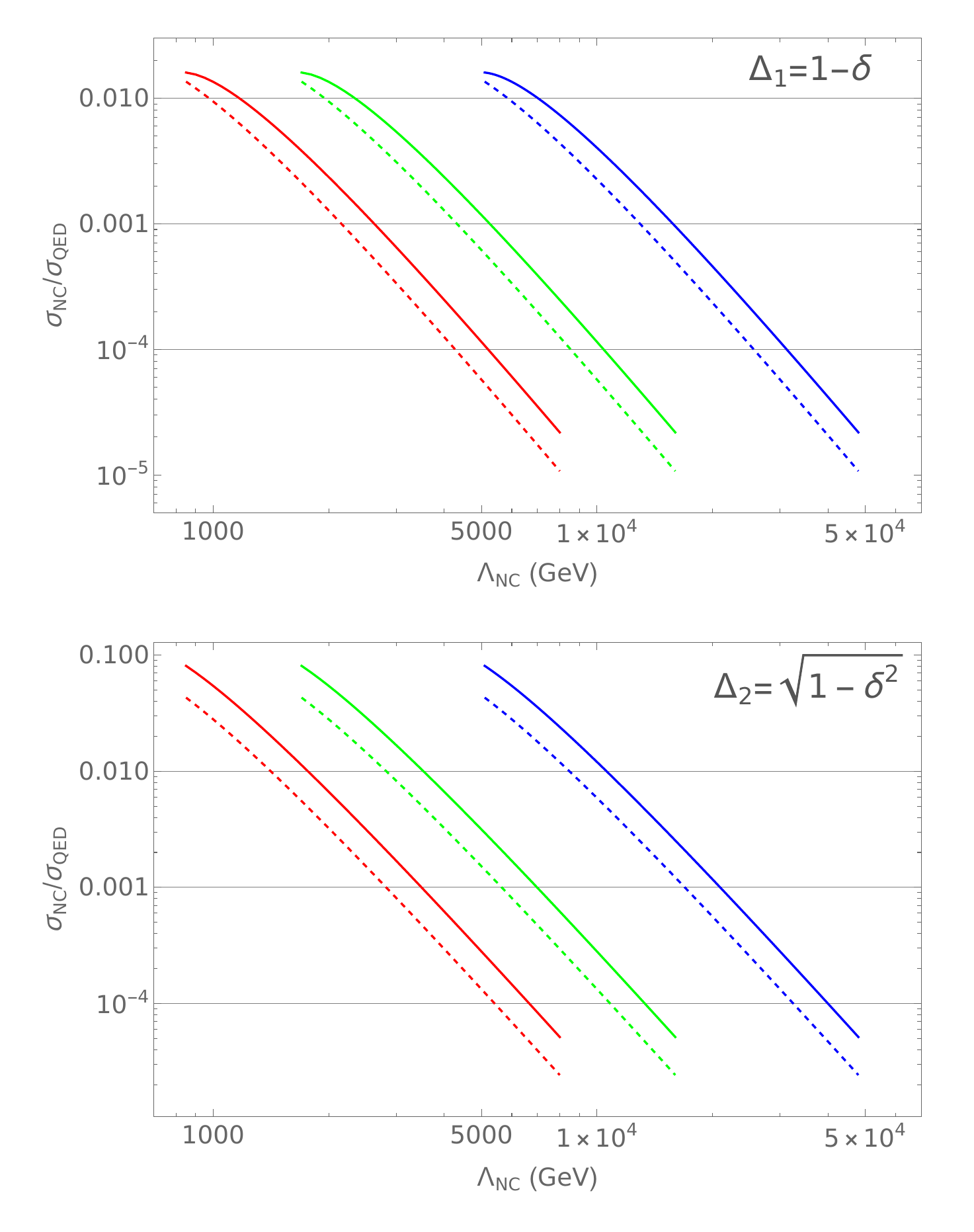} 
\end{center}
\caption{Relation between the integrated unpolarised cross sections ratio $\mathbb R$ and the NC scale $\Lambda_{\rm NC}$. Upper plots are calculated using $\Delta_1$, lower plots $\Delta_2$. Red lines are for $E^\gamma_{in}=200$ GeV, $E^e_{in}=250$ GeV collisions, green lines are for $E^\gamma_{in}=400$ GeV, $E^e_{in}=500$ GeV collisions, and blue lines are for $E^\gamma_{in}=1200$ GeV, $E^e_{in}=1500$ GeV collisions. Dashed lines are calculated using $x_0=0$ and solid lines are calculated using $x_0=0.5$. We marked 1\%, 0.1\%, and 0.01\% ratios with grid lines.}
\label{figure1}
\end{figure}

A fully consistent regularization of forward scattering singularity goes much beyond the scope of this article. On the other hand, we attempt to estimate here some phenomenological consequences of the forward scattering singularity using the idea of explicit regulator $\delta$. We assume that the actual regularization may be approximated by such cut-off for the integral cross sections. Since near the forward scattering singularity the NC differential cross section is suppressed by a prefactor $\delta^2$ any cut-off of the form $\delta^n$, where $n$ is a positive real number, shall make NC integral cross sections small enough when $\delta$ is sufficiently small. Therefore, two choices of cut-off $\Delta(\delta)$, the $\Delta_1(\delta)=1-\delta$ and the $\Delta_2(\delta)=\sqrt{1-\delta^2}$, were investigated. The resulting ratios are computed for the pure spacelike noncommutativity case (i), with $C^2= c_{13}^2+c_{23}^2\simeq 1$, and plotted as log-log scale in FIG.\ref{figure1}.

We notice that the integrated NC cross sections are fairly small. For the lowest energy scenario, the integrated NC cross section could get close to 1\% of the QED counterpart for $x_0=0$, or slightly over it for $x_0=0.5$ when $\Lambda_{\rm NC}$ is assumed to be 1 TeV and cut-off $\Delta_1$ is used. The cut-off $\Delta_2$ produces larger cross sections and allows 1\% ratio to be reached over 1 TeV for both $x_0$ values. The highest energy scenario allows 1\% ratio to be reached over 5 TeV for $\Delta_1$ and about 10 TeV for $\Delta_2$.

\section{Summary and conclusion}

In this work we compute in particularly noncommutative contributions to integral  unpolarised cross sections of the tree level $e\gamma\to e\gamma$ scattering in a Moyal $\theta$-exact NCQED model with only one generation of charged fermion. The $s/\Lambda_{\rm NC}^2$ dependent cut-offs were used in these computations to regulate the forward scattering singularity. We found that such cut-offs can, practically, suppress all contributions to inverse Compton scattering, since the regulated divergence gives a $\ln\delta$ enhancement while there is a $\delta^2$ suppression prefactor. We also considered the integral NC cross section over forward hemisphere ($x>0$) and forward $60^\circ$ cone ($x>0.5$) in a few $e\gamma$ collider scenarios and compared them with their commutative QED counterparts. Thank to the remnant of the singularity, these integral NC cross sections are relatively large with respect to their QED counterparts. We estimated that the NC integrated unpolarised cross section (\ref{TCrossSectC4}) reach about 1\% of the QED background for $\Lambda_{\rm NC}\gtrsim1$ TeV in $e\gamma\to e\gamma$ collisions, and is potentially realizable by next generations of linear colliders, which need to be high luminosity machines, with detectors constructed to cover relevant angular dependence also in the forward hemisphere and cone.

Considering pure QED contributions, note that radiative corrections \cite{Denner:1998nk,Lee:2021iid} increases integrated QED unpolarised cross section of $\sim 30\%$ at the CM energy $\sim1$ TeV, and thus decreases ratio $\mathbb R$ (\ref{R}), i.e. decreases trivially bound on the $\Lambda_{\rm NC}$ for $\lesssim 7\%$. Second, interference of the tree-level NCQED with one loop QED amplitude is suppressed with respect to the NC cross section (\ref{TCrossSectC4}) by more than two orders of magnitude ($\sim\alpha$), giving small contributions to the differential cross section. There is also an interference of tree-level NCQED with two loop QED amplitudes. 
At two loop order in QED emerges an almost energy independent forward scattering mechanism that involves two photons in the t–channel amplitude. 
However, that interference term ($\sim\alpha^4$) shall be suppressed with respect to (\ref{TCrossSectC4}) by $\alpha^2$. Inclusion of both such small contributions would certainly go beyond the scope of this paper at present stage of research for the spacetime noncommutativity. 
Finally, interference of the tree-level NCQED with the QED radiative corrections to unpolarised integral cross section for $\Lambda_{\rm NC}\gtrsim1$ TeV in $e\gamma\to e\gamma$ collisions is unlikely to be observed by the next generation of linear colliders.  

Our results are applicable to this Moyal NCQED model before and after a reversible SW map, thanks to an earlier proof that all $2\to2$ tree-level scattering amplitudes are invariant under reversible $\theta$-exact SW maps~\cite{Latas:2020nji}. It was also shown earlier that the same forward scattering divergence also exists in models with multiple generations of differently charged fermions~\cite{Trampetic:2022tij,Trampetic:2021awu}. So our considerations should apply to those models too.

Historically, potential NCQED corrections to the high energy Compton scattering on $e\gamma$ collider(s) were estimated quantitatively as changes of total cross sections (with unknown cut-off for the forward scattering singularity) and the Lorentz breaking transverse angular dependence~\cite{Godfrey:2001yy}. Our results in this work indicate that the unpolarised integral cross section of the forward regions could be more sensitive than the total cross section and allow us to probe  $\Lambda_{\rm NC}\gtrsim1$ TeV via $e\gamma\to e\gamma$ collisions. The Lorentz violating transverse angular ($\varphi$) variation of the differential cross section of $e\gamma\to e\gamma$ scattering should be a very sensitive parameter, thanks to the clean QED background. It is, however, worth to note that to observe the $\varphi$-variation cleanly would require that the NC parameter $\theta$ is unchanged with respect to the laboratory throughout the experiment. Otherwise the  $\varphi$-variation could be averaged out if (many) different $\theta$'s take place over the experiment time.  On the other hand, the forward scattering enhancement discussed here only vanishes if the NC parameters $\theta^{\mu\nu}$ are always so oriented that the factor $C$ in (\ref{TCrossSectC4}) vanishes. Therefore it should remain observable in the average when $\theta^{\mu\nu}$ is varying randomly over time. From this viewpoint the integral cross section of the forward regions may complement the transverse differential cross section variation when studying the possible signal from NCQED. Nevertheless, the NC effect in the form of integral cross sections from the forward orientations is still relatively hard to detect due to the QED background. Our estimations we made here are also intuitive, however they indicate that such an effect may be visible. 

Finally, it would be much more plausible to see a self-consistent regularization of the forward scattering singularity (and other IR divergences) in the NCQED in the future. We conclude that the forward scattering enhancement in the NC Compton scattering may also be a good place to search for the  phenomenological effects, if it could be properly regulated. 

\acknowledgments{J.T. would like to acknowledge support of Dieter L\"ust and Max-Planck-Institute for Physics, (Werner-Heisenberg-Institut), M\"unchen.}

\end{document}